# Onset of $AT_C$ superconductivity in $Ag_5Pb_2O_6$/CuO composite

D. Djurek

*Alessandro Volta Applied Ceramics (AVAC)*, Kesten brijeg 5. Remete, 10 000 Zagreb, Croatia

**Meissner type superconductivity extending up to 13 - 18 °C was confirmed in the composite material consisting of metallic particles $Ag_5Pb_2O_6$ dispersed in an insulating CuO matrix, and superconductivity mechanism is based upon inter particle interaction acting in the quantum tunnelling regime.**

In a recent work [1] author emphasized that several dozens convincing, but poorly reproducible, superconducting events at ambient temperature ($AT_c$) in fired and sintered mixtures $PbO–Ag_2O–CuO$ [2] and $PbO–Ag_2O–PbCO_3$ [3] were possibly associated with inter-grain interaction of Ag-defect Byström-Evers (BE) oxide $Ag_5Pb_2O_6$ [4] dispersed in the insulating matrices of CuO, $Pb_3O_4$ and $PbCO_3$.

Such a speculation was motivated by failure of many attempts to trace superconductivity in defect free $Ag_5Pb_2O_6$ obtained in the form of pellets fabricated from strongly pressed powders. Finally it was observed that loosely connected grains of Ag defect BE compound $Ag_{5-x}Pb_2O_6$ exhibit Meissner type SC properties up to 75-81 °C, giving rise to both low resistance states and diamagnetism [1]. An assumption of point contact superconductivity generated by inter grain interactions in quantum tunnelling regime appealed an attention for preparation of composites of defect free powdered $Ag_5Pb_2O_6$ mixed with the insulating oxides or carbonates. By recapitulation of first experiments [5] CuO appeared as a possible candidate. Although transition temperature $T_c$= 13–18 °C is somewhat lower than in Ag-defect samples, simplicity of preparation deserves to be presented here.

$Ag_5Pb_2O_6$ was prepared from the powders $PbO_2$ and $Ag_2O$ (Kemika, Zagreb) in corresponding molar ratio and mixed in a magnetic stirrer using ethanol as a mixing agent. After drying the mixture was fired for 24 hours at 310–325 °C and 130–140 bar $O_2$, which was followed by cooling to room temperature (RT), reground and repeated reheating for 24 hours under the same conditions. X-ray diffraction ($CuK_\alpha$) revealed the presence of $Pb_3O_4$ indicated by relative intensity 1.4 percent associated with the strongest diffraction hkl = 211.

The mean diameter of CuO and $Ag_5Pb_2O_6$ particles was determined by the adsorptive analysis and amounts 0.33 and 0.62 microns respectively.

Mixing of $Ag_5Pb_2O_6$ (5-2) with CuO powder in various weight to weight (w/w) proportions $\gamma = m_{5\text{-}2}/(m_{5\text{-}2}+m_{CuO})$ was again proceeded in ethanol and magnetic stirrer. Powders were dried and pressed in steel die to pellets 8 mm in diameter (inset of Fig.1), and series of preparations were grouped, each group was being indicated by the same volume of pellets. Constant volume among preparations within one group was ensured by the spacer (s) positioned between the male (m) and female (f) parts of the die. The thickness of the spacer in all these experiments was d=1.50 mm. The bare density of pellets within the same group was varied by different weights of powders imparted to the die hole. Gold wires 100 microns in diameter, arranged for four probe electric resistance measurements, were pressed together with pellet.

The pellets were then mounted on the ceramic sample holder protected by stainless steel tube [6], heated in 1 bar air atmosphere up to 322-326 °C and annealed for 24 hours.

Electric resistance was measured by the use of Keithley 6221 current source and Keithley 2700 multimeter.

The dependence of electric resistance of heat treated pellets (d = 1.50 mm) measured at 100 °C on the w/w ratio $\gamma$ is shown in Figure 1. It is evident an abrupt downturn of resistance in the range $\gamma = 0.40 - 0.55$ with a minimum which is followed by the near flat dependence up to $\gamma = 0.80$ and further decrease down to milliohm range for pure $A_5Pb_2O_6$ ($\gamma = 1$).

Measurements of ac susceptibility were performed on the pellets positioned in the assembly consisting of a primary coil and two compensated secondary coils. Out of balance signal was recorded by SR830 DSP lock-in amplifier. The measuring frequency was 322.80 Hz, and measuring field B = 0,82 gauss. The compensated coil measuring assembly was calibrated by the use of YBCO sample formed to the similar shape as the pellet, and it was assumed that such a reference exhibits full diamagnetism, that is, $1/4\pi$ emu·cm$^{-3}$. The temperature dependence of ac susceptibility for the pellet with parameters d=1.50 mm, $\gamma$=0.417 and $\rho$=4.67 g·cm$^{-3}$, recorded by cooling is shown in Figure 2. Onset temperature to SC state was 17 °C.

Figure 3 shows the temperature dependence of the resistance of pellet with parameters d=1.50 mm, $\gamma$=0.435, and $\rho$=5.97 g·cm$^{-3}$. The measuring current was 1 mA. The data are extended down to 5.3 °C given by the limited cooling capacity of the cold nitrogen gas stream supplied to the space between the stainless steel tube and the interior of the furnace. The resistance fluctuated in the range 6–8.5 °C, obviously as a result of temperature instability. The tube was then pulled out and plunged to liquid nitrogen bath. An additional control of the resistance was performed by lock-in amplifier supplying p-p current 94.5 mA at 213.5 Hz, and estimated resistivity at $LN_2$ temperature was less than $0.65 \cdot 10^{-8}$ ohm·cm.

Temperature dependence of the electric resistance of the pellet prepared for higher $\gamma=0.555$ is shown in Figure 4. There is evident a nearly linear dependence down to 25 °C, and value measured at $LN_2$ temperature was $R=0.504$ ohm, which is in good agreement with the value extrapolated from RT. No superconductivity was evidenced, either in resistance or in ac susceptibility measurements. The same holds for pellets prepared for lower concentrations of BE oxide in CuO, which is indicated by Figure 5 for value $\gamma=0.385$. The resistance diverges by cooling and is higher ha 1.5 kohm at $LN_2$ temperature. Like in High $T_c$ superconductors superconductivity is squeezed between the insulating and metallic states, when some exterior tuning parameter, in this case concentration of BE oxide in CuO matrix, is continuously varied. Similarly as in some other composites [7] like $Ni-SiO_2$ concentration of the metallic phase governs three regimes of electric properties. For low concentrations insulating or semiconducting behaviour dominates, and, by an increase of concentration, this regime is followed by quantum tunnelling state which is activated between the particles separated by close distance. Finally, for higher concentrations of the metallic particles percolation regime gives rise to various physical properties.

Cu–Pb–Ag–O system reported 1990 [2] has been the mixture of BE oxide, $Pb_3O_4$ and of major fraction CuO. There is a principal difference in approach when compared to data presented here. The outcome mixture was the result of fired powders CuO, PbO and $Ag_2O$, and despite of small fraction of conducting phase $Ag_5Pb_2O_6$ (molar 1.2 percents) the mixture was metallic, however not superconducting, except in small inhomogeneous regions. Approach presented here starts with BE oxide prepared independently and mixing with CuO was the next step. Difference is of fundamental significance, since molar BE fractions less than 30 percents in the mixture results in an insulator.

In conclusion, data reported in our previous work are confirmed, and the concept of the superconductivity radically differs from the commonly known superconductors when SC mechanism is activated within the crystal lattice, or in the case of High $T_c$ superconductors, within the grains. Mechanism reported here resides on the inter grain interactions favoured by quantum tunnelling.

In next papers various other oxides will be reported as insulating matrix carrying BE oxide, and the quality of samples will be certainly improved by introduction of mono-size particles with diameter reduced down to nanoscale dimensions.

**Figure captions:**

Fig. 1 Electric resistances of pellets (d = 1.50 mm) measured at 100 °C and plotted versus w/w ratio $\gamma = m_{5-2}/(m_{5-2}+m_{CuO})$. Density of pellets was the same $\rho$=4.67 g·cm$^{-3}$.

Fig. 2 Temperature dependence of ac magnetic susceptibility of the pellet $\gamma$=0.417.

Fig. 3 Temperature dependence of electric resistance of the pellet $\gamma$=0.435 and density $\rho$=5.97 g·cm$^{-3}$.

Fig. 4 Temperature dependence of electric resistance of the pellet $\gamma$=0.555.

Fig. 5 Temperature dependence of electric resistance of the pellet $\gamma$=0.385.

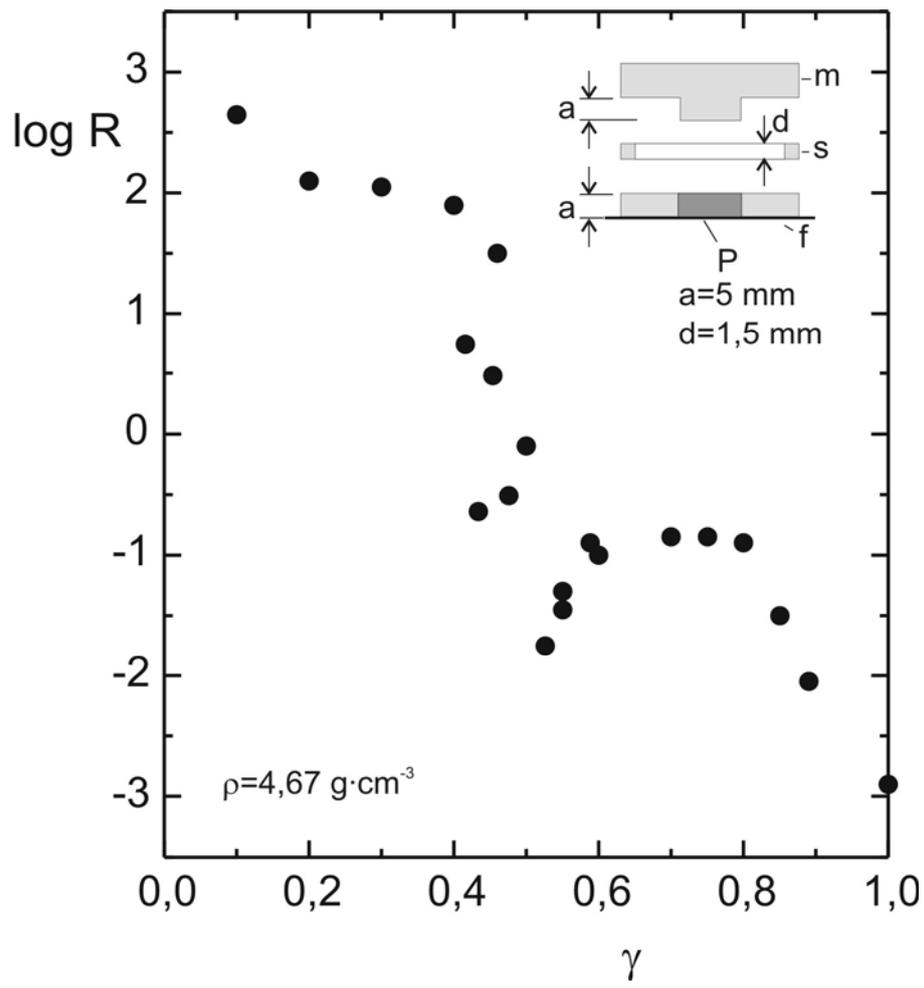

Figure 1.

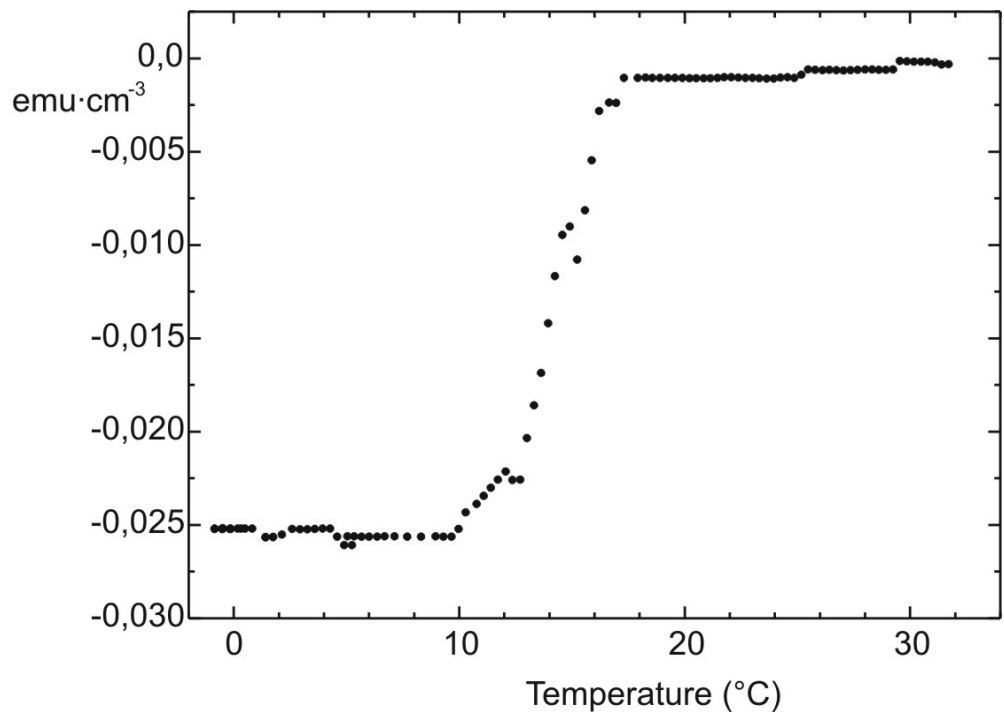

Figure 2.

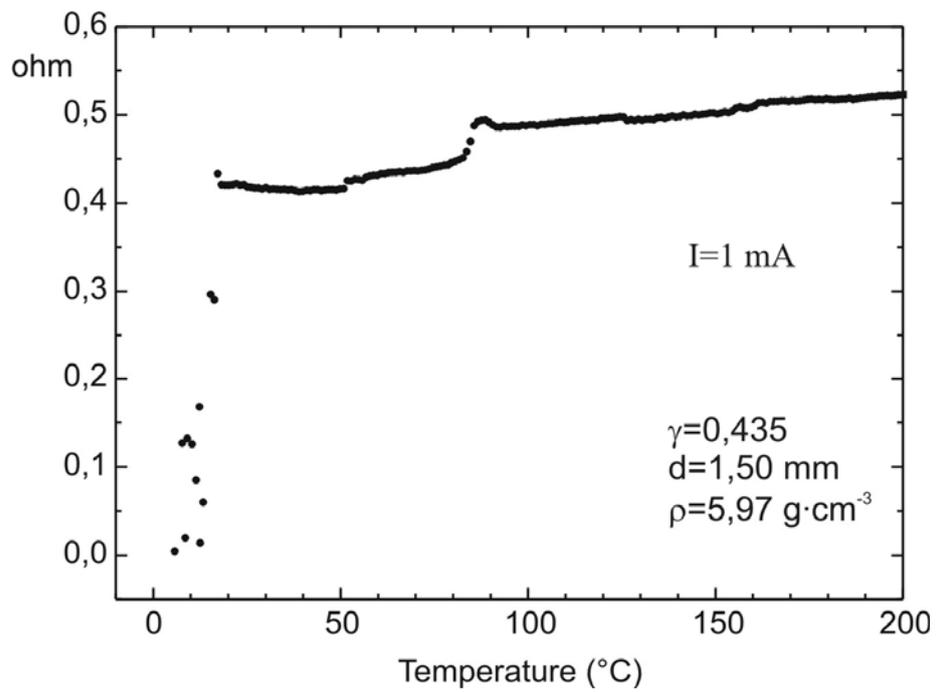

Figure 3.

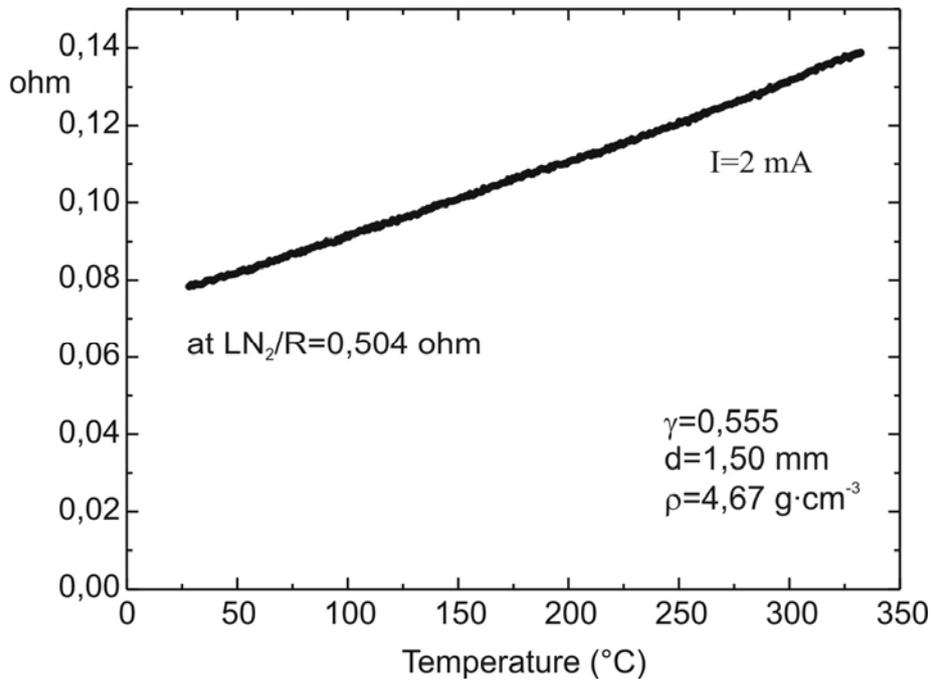

Figure 4

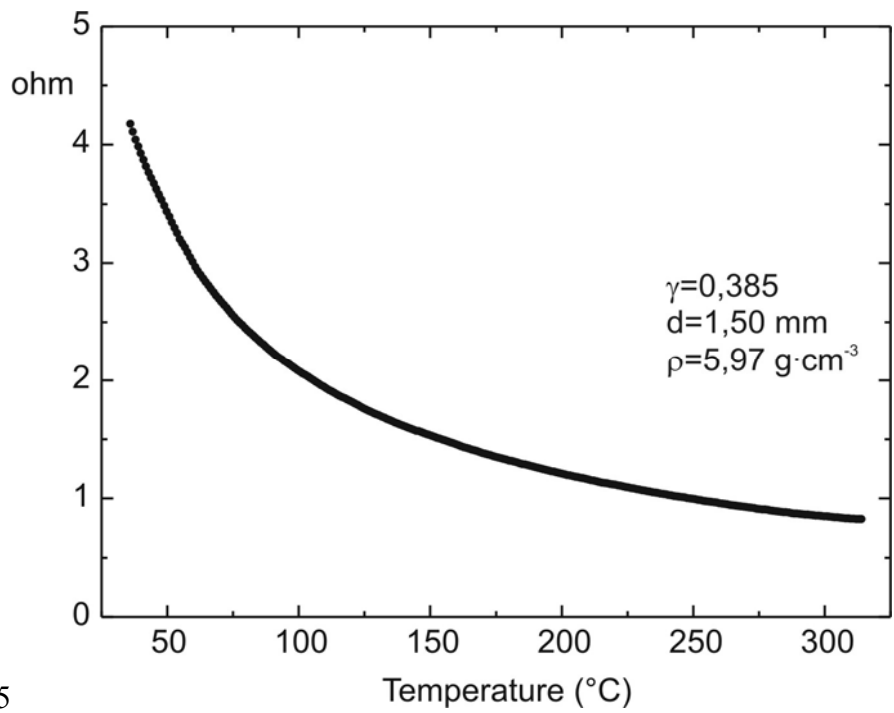

Figure 5